\documentclass[12pt]{article}
\usepackage[T1]{fontenc}
\usepackage{libertine}
\usepackage{tcolorbox}
\usepackage[top=1.2in, bottom=2in, left=0.8in, right=1in]{geometry}
\usepackage{float}
\usepackage{amsmath, amssymb, bm, tikz}
\usepackage{graphicx}
\usepackage{hyperref}
\usepackage{setspace}
\usepackage{parskip} 
\usepackage{xcolor} 
\usepackage{titlesec}

\usepackage{datetime} % For date formatting

\newdate{date}{30}{09}{2024} % Adjust to the appropriate date
\begin{document}
%\Large
\begin{center}
{\Large \textbf{A Breakthrough in Resonance Analysis for Complex Multilayered Structures in Small, High-Contrast Nonlinear Media with Kerr Effect}}\\[0.5cm]
{\large \textit{Taoufik Meklachi}}\\[0.2cm]
{\large Penn State Harrisburg}\\[0.2cm]
\displaydate{date}\\[1cm] % This displays the date formatted as specified
\end{center}
\large
\begin{abstract}
\noindent
Scattering resonances play a crucial role in understanding wave behavior in various physical systems. While significant progress has been made in analyzing resonances in high-contrast and nonlinear media, a general characterization of resonances in small, high-contrast nonlinear media with the Kerr effect, particularly in layered configurations, has remained an open problem. In this paper, we present a novel asymptotic approach that addresses this gap by providing an approximation for resonances in terms of material properties, geometry, and nonlinearity. Our results offer new insights into the dependence of resonances on these factors, particularly for complex multilayered structures, making this the first general characterization of its kind.
\end{abstract}
\section*{Introduction}

The study of scattering resonances in high-contrast media has attracted considerable attention due to its broad implications across various scientific and technological disciplines. These resonances, originating from the interaction of waves with material interfaces and inhomogeneities, underpin a wide range of phenomena, including light trapping in solar cells \cite{zhou2010trapping}, sound absorption in acoustic metamaterials \cite{cummer2016controlling}, and the manipulation of electromagnetic waves in advanced communication systems \cite{alu2007plasmonic}.

The ability to fabricate structures with intricate geometries and tailored optical properties has expanded significantly in recent years, thanks to advancements in material science and nanotechnology. However, analyzing wave propagation and resonance behavior in such complex systems often demands computationally intensive numerical simulations. Asymptotic methods, which offer approximate solutions in specific regimes, present a powerful alternative for gaining insights into these phenomena without relying on extensive numerical computations.

In this work, we delve into the development of an asymptotic approach to characterize scattering resonances in small, high-contrast nonlinear optical multilayered media exhibiting the Kerr effect. This nonlinearity, where the scattering resonances depend on both the refractive index and the intensity of the incident light, introduces an additional layer of complexity to the problem. Our objective is to derive an approximation for the resonances that illuminates their dependence on material properties, geometry, and the strength of the nonlinearity, particularly in the context of layered structures.

Prior research has explored resonances in high-contrast media using a variety of techniques. The first rigorous quantification of these resonances, in both linear and nonlinear regimes, was presented in \cite{Meklachi2018}, where a scaling approach and Lippmann-Schwinger integral solution to the Helmholtz equation were employed for a single volume. This work was subsequently extended in \cite{meklachi2022new} to cover small, high-contrast, multilayered linear media. Additionally, Ammari et al. \cite{ammari2018mathematical} established a mathematical framework for analyzing resonances in subwavelength resonator structures, primarily focusing on the linear regime. The influence of nonlinearity on resonance behavior has been investigated in several studies. For example, Moskow \cite{Moskow2015} examined nonlinear eigenvalue approximation for compact operators, providing a theoretical foundation for analyzing resonances in nonlinear systems. The work by Ammari et al. \cite{ammari2022layer} further extended these concepts to the context of high-contrast plasmonic media, employing layer potential techniques to approximate resonances.

Moreover, the study of resonances in nonlinear optical media has been a subject of active research. The Kerr effect, in particular, has been extensively investigated due to its role in phenomena like self-focusing, soliton formation, and optical bistability \cite{boyd2020nonlinear}. Theoretical and numerical studies have explored the impact of the Kerr nonlinearity on resonance frequencies and mode profiles in various configurations \cite{poutrina2010nonlinear, suchkov2000resonant}.

While these previous works have significantly advanced our understanding of resonances in high-contrast and nonlinear media, a general characterization of resonances in small, high-contrast nonlinear media with Kerr effect, especially in layered configurations, remains an open problem. Our work aims to bridge this gap by providing an asymptotic approximation that encapsulates the essential physics of the problem. We build upon the foundations laid by previous research, extending them to incorporate the nonlinear Kerr effect and providing a more comprehensive understanding of resonance phenomena in these complex systems.

The potential applications of our research are vast and diverse. In the domain of photonics, our findings could facilitate the design of novel metamaterials with exotic optical properties, enabling advancements in optical sensing, light manipulation, and energy harvesting \cite{soukoulis2011past}. In the realm of acoustics, the ability to precisely control resonances in small structures could lead to the development of acoustic metamaterials with superior sound control capabilities, impacting noise reduction and architectural acoustics \cite{ma2014acoustic}. Furthermore, our work could have ramifications for medical imaging \cite{medimage} , where resonant nanoparticles could function as contrast agents or enable targeted drug delivery \cite{kircher2012resonant}. In electronics and telecommunications, a deeper understanding of resonances could lead to the design of smaller, more efficient antennas and improved electromagnetic shielding \cite{alitalo2015graphene}. Finally, our research could contribute to the development of more efficient energy harvesting and conversion devices, such as solar cells and thermoelectric devices \cite{atwater2010plasmonics}.

This work is an expansion of the results in \cite{Meklachi2018} and \cite{meklachi2022new}. So section 1.1 will present the major results and solution methods for one small high contrast nonlinear medium discussed in \cite{Meklachi2018}. The approach used in \cite{meklachi2022new} is key to the expansion to multilayered media, as we show in the subsequent sections. We believe that this general characterization of resonances in small, multilayered, high-contrast nonlinear optical media with Kerr effect is novel and has the potential to significantly impact a wide range of scientific and technological fields.

\section{Resonances in Small, Doublelayered,
High-Contrast, Nonlinear Scatterers with Arbitrary Geometry}
\subsection{The original work}
In this section we present the results in \cite{Meklachi2018}, as we use aspects of the solution method to compute the new formulae. The reader can refer to \cite{Meklachi2018} for more details and examples. 

Consider a small volume high contrast nonlinear medium exhibiting Kerr effect of arbitrary geometry. The governing equation is
 \[
	\Delta u+k^2(1+\eta(x))u+k^2\beta(x)|u|^2u=0 \ 
	\]
	
	subject to Sommerfeld radiation condition at infinity.
	
	\noindent where $k$ is the wave number, $\eta$ is the medium susceptibility coefficient, and $u$ is the scalar field. Here, $\lambda=k^2$ is the spectral parameter.
	
 Lippmann-Schwinger Integral solution is given by:
	\begin{align*}\label{eq:lip2}
		u(x) &= u_i(x) + k^2\int_{hB}\frac{\eta_0(y)}{h^2}G(x,y)u(y)dy \notag \\
		&\quad + k^2\int_{hB}\frac{\beta_0(y)}{h^2}G(x,y)|u(y)|^2u(y)dy \ ,
	\end{align*}
	
	\noindent where $u_i$ is the incident field and $G$ in 3D is defined as:
	
	\[
	G(x,y)=\frac{1}{4\pi}\frac{\exp({i\sqrt{\lambda}|x-y|)}}{|x-y|}
	\]
	
	\vspace{1cm}
	
	Setting $u_i=0$, we obtain the integral eigenvalue problem that is nonlinear in both $\lambda$ and $u$: 
	
	\[
	\lambda T(\lambda)u=u
	\]  
	where
	
	\[
	T(\lambda)(u)(x)=\frac{1}{4\pi}\int_{hB}  \eta(y)\frac{\exp({i\sqrt{\lambda}|x-y|)}}{|x-y|}u(y)dy \qquad
	\]
	
	\[
	+\frac{1}{4\pi}\int_{hB}\beta(y)\frac{\exp({i\sqrt{\lambda}|x-y|)}}{|x-y|}|u(y)|^2u(y)dy
	\]
	
	\vspace{1cm}
	By scaling the spatial variables and the media properties as follows: 
	\[ 
	\eta=\chi_{hB}\frac{\eta_0}{h^2}, \quad \beta=\chi_{hB}\frac{\beta_0}{h^2},\quad x=h\tilde{x}, \quad y=h\tilde{y}
	\]
	we obtain:
	\[
	\lambda_h T_h(\lambda_h)u_h=u_h
	\]  
	where
	\begin{equation}\label{Th}
	T_h(\lambda)(u)(\tilde{x})=\frac{1}{4\pi}\int_B  \eta_0(h\tilde{y})\frac{\exp({i\sqrt{\lambda}h|\tilde{x}-\tilde{y}|)}}{|\tilde{x}-\tilde{y}|}u(\tilde{y})d\tilde{y}
	\end{equation}
	\[
	+\frac{1}{4\pi}\int_B  \beta_0(h\tilde{y})\frac{\exp({i\sqrt{\lambda}h|\tilde{x}-\tilde{y}|)}}{|\tilde{x}-\tilde{y}|}u(\tilde{y})|u(\tilde{y})|^2d\tilde{y}
	\]
 
The associated limiting eigenvalue problem, as $h \rightarrow 0$ is:
	
	\[
	\lambda_0	T_0(\lambda_0)u_0=u_0 \ ,
	\]
	where
	\[
	T_0(u)(\tilde{x})=\frac{1}{4\pi}\int_B  \eta_0(0)\frac{u(\tilde{y})}{|\tilde{x}-\tilde{y}|}d\tilde{y}+\frac{1}{4\pi}\int_B  \beta_0(0)\frac{u(\tilde{y})}{|\tilde{x}-\tilde{y}|}|u(\tilde{y})|^2d\tilde{y} \]
	
	Now that $T_h$ and $T_0$ are well defined, we can compute the asymptotic formula by executing the inner product in the formula 
 
	\[
	\lambda_h=\lambda_0+{\lambda_0}^2\left\langle(T_0-T_h(\lambda_0))u_0,u_0\right\rangle+\mathcal{O}(h^2)
	\]
	$T_0$ is a self-adjoint compact operator with positive eigenvalues \cite{Moskow2015}. Performing Taylor exapnsion on the functions $$h \longmapsto\eta_0(h\tilde{y})\frac{\exp({i\sqrt{\lambda}h|\tilde{x}-\tilde{y}|)}}{|\tilde{x}-\tilde{y}|}
\quad  \text{and} \quad h \longmapsto\beta_0(h\tilde{y})\frac{\exp({i\sqrt{\lambda}h|\tilde{x}-\tilde{y}|)}}{|\tilde{x}-\tilde{y}|}
$$
gives a first order approximation of the resonances as follows. See \cite{Meklachi2018} for more details.	
Let $$ U_0=\int_B u_0(x)dx, \quad \text{and  } \quad u_h=u_0+u_1h+o(h)$$ 
   The value $\lambda_1$ in $\lambda_h=\lambda_0+\lambda_1 h+\mathcal{O}(h^2)$ is given by:
	
	\begin{multline}\label{correction}
		-4\pi\lambda_1=2\beta_0(0)\lambda_0^2\int_B\int_B\frac{u_0(y)Re(u_1(y))}{|x-y|}u_0(x)dydx \\ 
		+\lambda_0^2\int_B\int_B\frac{\nabla\eta_0 (0) \cdot y u_0(y)}{|x-y|}u_0(x)dydx +i\lambda_0^{\frac{5}{2}}\eta_0(0)U^2_0 \\ 
		+\lambda_0^2\int_B\int_B\frac{\nabla\beta_0 (0) \cdot y u^3_0(y)}{|x-y|}u_0(x)dydx+i\lambda_0^{\frac{5}{2}}\beta_0(0)U_0\int_Bu^3_0(y)dy
	\end{multline}
	\begin{multline*}
		\text{Re($u_1$) obeys: \qquad} 
		4\pi Re(u_1(x))=\\ 
		\lambda_0\int_B\frac{\nabla\eta_0\cdot yu_0+\nabla\beta_0\cdot yu_0^3}{|x-y|}dy \\
		+\lambda_0\int_B\frac{Re(u_1)}{|x-y|}(\eta_0+3\beta_0u_0^2)dy \\
		-\frac{\lambda_0}{4\pi}u_0(x)\bigg(2\beta_0(0)\int_B\int_B\frac{u_0(y)Re(u_1(y))}{|x-y|}u_0(x)dydx \\
		+\int_B\int_B\frac{\nabla\eta_0 (0) \cdot y  u_0(y)}{|x-y|}u_0(x)dydx +\int_B\int_B\frac{\nabla\beta_0 (0) \cdot y u^3_0(y)}{|x-y|}dydx\bigg)
	\end{multline*}
 In practice, we will show that few terms in formula \eqref{correction} are zero. For example, when $\eta_0$ and $\beta_0$ are constants, corollary 3.1 shows that:  
  \begin{align}\label{eq:lambda13}
 -4\pi\lambda_1=2\beta_0\lambda_0^2\int_B\int_B\frac{u_0(y)Re(u_1)}{|x-y|}u_0(x)dydx+
 &i\lambda_0^{\frac{5}{2}}\eta_0U^2_0+i\lambda_0^{\frac{5}{2}}\beta_0U_0U_1
 \end{align}
 where
\begin{equation*}
U_0=\int_B u_0(x)dx \text{  ,   }   U_1=\int_Bu_0(x)^3dx  
\end{equation*}
and  $Re(u_1)$ solves the integral equation: 
\begin{equation}\label{eq:re_u1}
\begin{split}
4\pi Re(u_1(x))=&\lambda_0\int_B\frac{Re(u_1)}{|x-y|}(\eta_0+3\beta_0u_0^2)dy\\
&-\frac{\lambda_0}{4\pi}2\beta_0u_0(x)\int_B\int_B\frac{u_0(y)Re(u_1)}{|x-y|}u_0(x)dydx.
\end{split}
\end{equation} 
 Furthermore, in the specific case of a spherical scatterer, we found that $R(u_1)$ vanishes, i.e., $R(u_1) = 0$. 

While the general derivation presented in equation \eqref{correction} may appear complex and potentially challenging to implement due to the non-constant nature of $\eta$ and $\beta$, and the arbitrary geometry involved, it often simplifies significantly in practical physical scenarios. This apparent inconvenience actually proves to be a major advantage, a point this paper will thoroughly illustrate.
\subsection{Case 1: Single-Layer Kerr Effect}
	Consider the scattering problem of high contrast small volume, $hB$. The small scatterer has two concentric inner and outer layers, with the latter exhibiting nonlinear Kerr effect. 
	\[
	hB = hB_{\text{in}} \cup hB_{\text{out}}
	\]
	\[
	\eta(x) = \chi_{hB_{\text{in}}}(x)\eta_{\text{in}} + \chi_{hB_{\text{out}}}(x)\eta_{\text{out}} = \chi_{hB_{\text{in}}}\frac{\eta_{\text{in}}}{h^2} + \chi_{hB_{\text{out}}}\frac{\eta_{\text{out}}}{h^2}
	\]
	\[
	\beta(x) = \chi_{hB_{out}}\frac{\beta_0}{h^2}
	\]
	for constant $\eta_0$ and $\beta_0$.
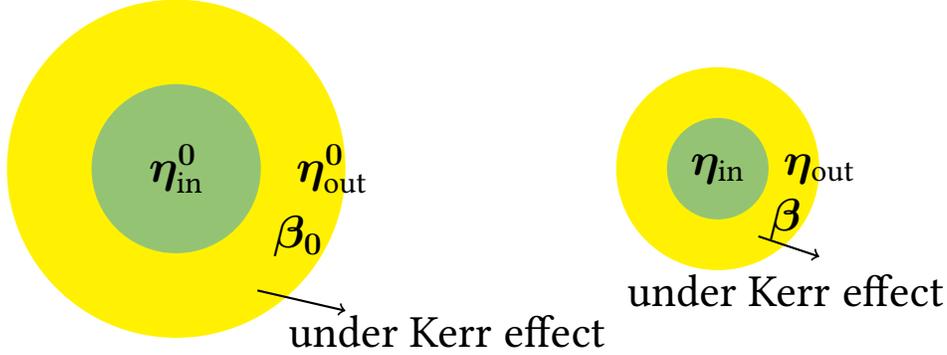
\begin{center}
    \begin{tikzpicture}[scale=0.9] % Scaling the whole picture up
        % First (left) Outer circle (yellow) - made larger
        \fill[yellow] (0,0) circle (2.5cm); % Outer yellow circle made slightly larger (2.5cm)
        
        % First (left) Inner circle (pistachio green) - adjusted accordingly
        \definecolor{pistachio}{rgb}{0.58, 0.77, 0.45} % Define pistachio green color
        \fill[pistachio] (0,0) circle (1.25cm); % Inner pistachio green circle made slightly larger (1.25cm)
        
        % Outer label with eta_0 (left circle)
        \node at (2.3, 0) {\Large $\boldsymbol{\eta^0_{\text{out}}}$}; % Outer label shifted slightly for larger circle in bold and larger size
        \node at (1.8, -1) {\Large $\boldsymbol{\beta_0}$}; % Adjusted position for beta_0 label in bold and larger size
        
        % Inner label (left circle)
        \node at (0,0) {\Large $\boldsymbol{\eta^0_{\text{in}}}$}; % Inner label in bold and larger size
        
        % Arrow indicating Kerr effect (left circle) - adjusted position
        \draw[->, thick] (1.2, -1.8) -- (2.5, -2.1); % Adjusted arrow for larger circle
        \node at (4, -2.4) {\Large under Kerr effect}; % Larger text for under Kerr effect
        
        % Second (right) Outer circle (yellow)
        \fill[yellow] (8,0) circle (1.5cm); % Smaller outer yellow circle on the right
        
        % Second (right) Inner circle (pistachio green)
        \fill[pistachio] (8,0) circle (0.75cm); % Smaller inner pistachio green circle on the right
        
        % Outer label with eta (right circle)
        \node at (9.5, 0) {\Large $\boldsymbol{\eta_{\text{out}}}$}; % Smaller outer label shifted slightly in bold and larger size
        \node at (9, -0.75) {\Large $\boldsymbol{\beta}$}; % Smaller beta label in bold and larger size
        
        % Inner label (right circle)
        \node at (8,0) {\Large $\boldsymbol{\eta_{\text{in}}}$}; % Smaller inner label in bold and larger size
        
        % Arrow indicating Kerr effect (right circle)
        \draw[->, thick] (8.6, -1) -- (9.5, -1.3);
        \node at (9, -1.8) {\Large under Kerr effect}; % Larger text for under Kerr effect
        
    \end{tikzpicture}
\end{center}
	Within the context of this new configuration, the operators $T_h$, previously defined by \eqref{Th}, can be expressed as:
\begin{align*}
	T_h(\lambda)(u)(x) &= \frac{\eta^{0}_{in}}{4\pi}\int_{B_{in}}\frac{\exp({i\sqrt{\lambda}h|x-y|)}}{|x-y|}u(y)dy \\
	&\quad + \frac{\eta^{0}_{out}}{4\pi}\int_{B_{out}}\frac{\exp({i\sqrt{\lambda}h|x-y|)}}{|x-y|}u(y)dy \\
	&\quad + \frac{\beta_0}{4\pi}\int_{B_{out}} \frac{\exp({i\sqrt{\lambda}h|x-y|)}}{|x-y|}u(y)|u(y)|^2dy
\end{align*}
Hence we define the limiting operator, $T_0$, as $h\rightarrow 0$ as:
\begin{align*}
	T_0(\lambda)(u)(x) &= \frac{\eta^{0}_{in}}{4\pi}\int_{B_{in}}\frac{u(y)}{|x-y|}dy \\
	&\quad + \frac{\eta^{0}_{out}}{4\pi}\int_{B_{out}} \frac{u(y)}{|x-y|}dy \\
	&\quad + \frac{\beta_0}{4\pi}\int_{B_{out}} \frac{u(y)}{|x-y|}|u(y)|^2dy
\end{align*}
As it was introduced in section 1.1,  we are solving for the resonances, $\lambda_h$, the eigenvalue problem $$\lambda_hT_h u_h=u_h$$ via the limiting eigenvalue problem $$\lambda_0T_0 u_0=u_0$$
In other words, we first solve the easier latter equation for $u_0$ and $\lambda_0$.
Now let us expand $T_h$ by performing Taylor expansion of $h \longmapsto \exp(i\sqrt{\lambda}h|x-y|)$:

\begin{align*}
	&T_h(\lambda)(u)(x) = T_0(\lambda)(u)(x) + \frac{i\sqrt{\lambda}h}{4\pi} \left( \eta^{0}_{in} \int_{B_{in}} u(y) dy \right. \\
	& \left. + \eta^{0}_{out} \int_{B_{out}} u(y) dy + \beta_0 \int_{B_{out}} u(y) |u(y)|^2 dy \right) + \mathcal{O}(h^2)
\end{align*}
Evaluating at $u_0$ and $\lambda_0$ we obtain:
\begin{align*}
&(T_0 - T_h(\lambda_0))u_0 = -\frac{i\sqrt{\lambda_0}h}{4\pi} \left( \eta^{0}_{in} \int_{B_{in}} u_0(y) dy \right. \\
&\left. + \eta^{0}_{out} \int_{B_{out}} u_0(y) dy + \beta_0 \int_{B_{out}}u^3_0(y)dy \right) + \mathcal{O}(h^2)
\end{align*}
or
\[
(T_0 - T_h(\lambda_0))u_0 = -\frac{i\sqrt{\lambda_0}h}{4\pi} \left( \eta^{0}_{in} U_{in}
 + \eta^{0}_{out} U_{out} + \beta_0 U_{out\beta} \right) + \mathcal{O}(h^2)
\]
where
$$U_{in}=\int_{B_{in}} u_0(y) dy,\quad U_{out}=\int_{B_{out}} u_0(y) dy $$
and $$U_{out\beta}=\int_{B_{out}} u^3_0(y) dy$$

 At this point, we are ready to compute $\lambda_1$ by executing the expansion asymptotic formula:
\begin{equation}\label{main}
\lambda_h = \lambda_0 + \lambda_0^2 \langle (T_0 - T_h(\lambda_0))u_0, u_0 \rangle + \mathcal{O}(h^2)
\end{equation}
Let $$\lambda_h=\lambda_0+\lambda_1 h+ \mathcal{O}(h^2)$$
	It follows that the correction $\lambda_1$ is 
	\begin{center}
    \begin{tcolorbox}[colback=lightgray!20, colframe=blue!50, boxrule=0.75mm, sharp corners, width=0.7\textwidth]
        \begin{equation*}\label{case1}
        \boldsymbol{\lambda_1 = -\frac{i\lambda^{\frac{5}{2}}_0}{4\pi} \left( \eta^{0}_{in} U_{in}
        + \eta^{0}_{out} U_{out} + \beta_0 U_{out\beta} \right)U_0}
        \end{equation*}
    \end{tcolorbox}
\end{center}
 where $$U_0=\int_B u_0(y) dy$$
 \newpage

  \subsection{Case 2: Dual-Layer Kerr Effect}
 This configuration offers greater flexibility in designing the 3D scatterer, allowing for scenarios with either distinct or identical susceptibility coefficients ($\eta_{\text{in}} \neq \eta_{\text{out}}$ or $\eta_{\text{in}} = \eta_{\text{out}}$) and Kerr coefficients ($\beta_{\text{in}} \neq \beta_{\text{out}}$ or $\beta_{\text{in}} = \beta_{\text{out}}$). All coefficients are assumed to be constant.
  \vspace{0.7cm}
  \[ 
	\beta(x) = \chi_{hB_{in}}\frac{\beta^0_{in}}{h^2}+
\chi_{hB_{out}}\frac{\beta^0_{out}}{h^2}
	\]
   \vspace{0.4cm}

\begin{center}
    \begin{tikzpicture}[scale=1.1] % Scaling the whole picture
        % First (left) Outer circle (yellow) - made larger
        \fill[yellow] (0,0) circle (2.5cm); % Outer yellow circle made slightly larger (2.5cm)
        
        % First (left) Inner circle (pistachio green) - adjusted accordingly
        \definecolor{pistachio}{rgb}{0.58, 0.77, 0.45} % Define pistachio green color
        \fill[pistachio] (0,0) circle (1.25cm); % Inner pistachio green circle made slightly larger (1.25cm)
        
        % Outer label with eta_0 (left circle)
        \node at (2.3, 0) {\huge $\mathbf{\eta^0_{\text{out}}}$}; % Outer label with larger font
        
        % Inner labels (left circle)
        \node at (0,0.5) {\huge $\mathbf{\eta^0_{\text{in}}}$}; % Inner label with larger font
        \node at (0,-0.5) {\huge $\mathbf{\beta^0_{\text{in}}}$}; % Inner circle label for beta_1 with larger font
        
        % Outer circle label for beta_2
        \node at (1.5,1.5) {\huge $\mathbf{\beta^0_{\text{out}}}$}; % Outer circle label for beta_2 with larger font
        
        % Second (right) Outer circle (yellow)
        \fill[yellow] (8,0) circle (1.5cm); % Smaller outer yellow circle on the right
        
        % Second (right) Inner circle (pistachio green)
        \fill[pistachio] (8,0) circle (0.75cm); % Smaller inner pistachio green circle on the right
        
        % Outer label with eta (right circle)
        \node at (9.5, 0) {\huge $\mathbf{\eta_{\text{out}}}$}; % Smaller outer label with larger font
        
        % Inner labels (right circle)
        \node at (8,0.3) {\huge $\mathbf{\eta_{\text{in}}}$}; % Smaller inner label with larger font
        \node at (8,-0.3) {\huge $\mathbf{\beta_{\text{in}}}$}; % Inner circle label for beta_1 with larger font
        
        % Outer circle label for beta_2
        \node at (8.9, 0.75) {\huge $\mathbf{\beta_{\text{out}}}$}; % Outer circle label for beta_2 with larger font
    \end{tikzpicture}
\end{center}
	In this case $T_h$ and $T_0$ write:
\begin{align*}
	T_h(\lambda)(u)(x) &= \frac{\eta^{0}_{in}}{4\pi}\int_{B_{in}}\frac{\exp({i\sqrt{\lambda}h|x-y|)}}{|x-y|}u(y)dy \\
	&\quad + \frac{\eta^{0}_{out}}{4\pi}\int_{B_{out}}\frac{\exp({i\sqrt{\lambda}h|x-y|)}}{|x-y|}u(y)dy \\
	&\quad + \frac{\beta^0_{in}}{4\pi}\int_{B_{in}} \frac{\exp({i\sqrt{\lambda}h|x-y|)}}{|x-y|}u(y)|u(y)|^2dy\\
	&\quad + \frac{\beta^0_{out}}{4\pi}\int_{B_{out}} \frac{\exp({i\sqrt{\lambda}h|x-y|)}}{|x-y|}u(y)|u(y)|^2dy\end{align*}
and
\begin{align*}
	T_0(\lambda)(u)(x) &= \frac{\eta^{0}_{in}}{4\pi}\int_{B_{in}}\frac{u(y)}{|x-y|}dy \\
	&\quad + \frac{\eta^{0}_{out}}{4\pi}\int_{B_{out}} \frac{u(y)}{|x-y|}dy \\
	&\quad + \frac{\beta^0_{in}}{4\pi}\int_{B_{in}} \frac{u(y)}{|x-y|}|u(y)|^2dy
 \\
	&\quad + \frac{\beta^0_{out}}{4\pi}\int_{B_{out}} \frac{u(y)}{|x-y|}|u(y)|^2dy
\end{align*}
Similar to case 1, we perform a Taylor expansion of $h \longmapsto \exp(i\sqrt{\lambda}h|x-y|)$, yielding:

\begin{align*}
&T_h(\lambda)(u)(x) = T_0(\lambda)(u)(x) + \frac{i\sqrt{\lambda}h}{4\pi} \left( \eta^{0}_{in} \int_{B_{in}} u(y) dy \right. \\
&\quad \left. + \eta^{0}_{out} \int_{B_{out}} u(y) dy + \beta^{0}_{in} \int_{B_{in}} u(y) |u(y)|^2 dy + \beta^{0}_{out} \int_{B_{out}} u(y) |u(y)|^2 dy \right) +\mathcal{O}(h^2)
\end{align*}
Subsequently, the correction $\lambda_1$ is derived by evaluating the inner product in equation \eqref{main}.
 \begin{center}
    \begin{tcolorbox}[colback=lightgray!20, colframe=blue!50, boxrule=0.75mm, sharp corners, width=0.9\textwidth]
        \begin{equation*}\label{case2}
        \boldsymbol{
            \lambda_1 = -\frac{i\lambda^{\frac{5}{2}}_0}{4\pi} \left( \eta^{0}_{in} U_{in}
            + \eta^{0}_{out} U_{out} + \beta^0_{in} U_{in\beta} + \beta^0_{out} U_{out\beta}\right)U_0}
                \end{equation*}

    \end{tcolorbox}
\end{center}
 where
 $$U_{in\beta}=\int_{B_{in}} u^3_0(y) dy, \quad U_{out\beta}=\int_{B_{out}} u^3_0(y) dy$$
 $$U_{in}=\int_{B_{in}} u_0(y) dy,\quad U_{out}=\int_{B_{out}} u_0(y) dy, \,
\text{and} \quad U_0=\int_{B} u_0(y) dy$$
Note that we recover $\lambda_1$ in case 1 by setting $\beta^0_{in}=0$. Furthermore, setting $\beta^0_{out}=0$ provides the formula for $\lambda_1$ when only the inner layer exhibits the Kerr effect.
\section{Resonances in Small, Multilayered,
High-Contrast, Nonlinear Scatterers with Arbitrary Geometry}
 
   \vspace{0.8cm}
   	Suppose $B_h$ is composite and optically inhomogenous with $n$ layers of concentric high contrast media ${\{B_i}\}_{1\leq i\leq n}$, such that
	\[hB=\cup_{i=1}^{i=n} hB_i, \]
	\[\eta(x)=\sum_{i=1}^{i=n}\chi_{hB_i}(x)\eta^i =\sum_{i=1}^{i=n}\chi_{hB_i}(x)\frac{\eta_0^i}{h^2},\]
 \[\beta(x)=\sum_{i=1}^{i=n}\chi_{hB_i}(x)\beta^i =\sum_{i=1}^{i=n}\chi_{hB_i}(x)\frac{\beta_0^i}{h^2},\]
	\[\eta_0(x)=\sum_{i=1}^{i=n}\chi_{B_i}(x)\eta_0^i, \quad 
 \beta_0(x)=\sum_{i=1}^{i=n}\chi_{B_i}(x)\beta_0^i\]
 Similar to the previous cases, the operators $T_h$ and $T_0$ write:
 \begin{multline}
		T_h(u)({x})(\lambda)=\sum_{k=1}^{k=n}\frac{\eta_0^k}{4\pi}\int_{B_k}\frac{\exp{(i\sqrt{\lambda }h|x-y|})}{|x-y|}u(y)dy
  \\
  +\frac{\beta_0^k}{4\pi}\int_{B_k}\frac{\exp{(i\sqrt{\lambda }h|x-y|})}{|x-y|}u(y)|u(y)|^2dy
\end{multline}	\begin{equation}
		T_0(u)({x})=\sum_{k=1}^{k=n}\frac{\eta_0^k}{4\pi}\int_{B_k}\frac{1}{|x-y|}u(y)dy+\frac{\beta_0^k}{4\pi}\int_{B_k}\frac{1}{|x-y|}u(y)|u(y)|^2dy
  \end{equation}
  We perform a Taylor expansion of $h \longmapsto \exp(i\sqrt{\lambda}h|x-y|)$ again, yielding:

\begin{equation*}
T_h(\lambda)(u)(x) = T_0(\lambda)(u)(x) + \frac{i\sqrt{\lambda}h}{4\pi} \left( \sum_{k=1}^{n} \eta_0^k \int_{B_{k}} u(y) dy + \beta_0^k \int_{B_{k}} u(y) |u(y)|^2 dy + \right) +\mathcal{O}(h^2)
\end{equation*}
We can now evaluate \eqref{main} at $u_0$ and $\lambda_0$, which gives:
\begin{center}
    \begin{tcolorbox}[colback=yellow!20!orange!20, colframe=yellow!50!orange!80, boxrule=1mm, sharp corners, width=\textwidth]
        \[
        \boldsymbol{
            \lambda_h = \lambda_0 - i\frac{{\lambda_0}^\frac{5}{2}}{4\pi} U_0
            \left[  \sum_{k=1}^{n} \eta_0^k U_k + \beta_0^k U^\beta_k \right] h + \mathcal{O}(h^2)
        }
        \]
    \end{tcolorbox}
\end{center}

		where  \[U_0=\int_{B}u_0(x)dx,\]
	and
	$$U_k=\int_{B_k}u_0(x)dx, \quad U^\beta_k=\int_{B_k}u^3_0(x)dx, \quad 1\leq k\leq n.$$
In the future, we aim to investigate the higher-order terms, such as the second-order component of $\lambda_h$, to gain better control over the physical parameters contributing to the scattering phenomenon in small, high-contrast nonlinear media. A similar investigation was conducted for linear media in \cite{Meklachi2022}.
  \newpage

\bibliography{references}
\bibliographystyle{siam}
\end{document}